\documentclass[10pt,a4wide,twoside]{article}

\usepackage{epsfig,graphicx,fancyhdr}
\usepackage{natbib}
\usepackage{amssymb} 
\usepackage{amsmath}
\begin{document}
\sf

\title{Effects of resolution and helium abundance in A star surface convection 
simulations\thanks{preprint of version submitted to CoAst}}
\markboth{Effects of resolution and helium abundance in A star surface convection 
simulations}{F.\,Kupka, J.\,Ballot, and H.J.\,Muthsam}
\author{F.\,Kupka \\ 
                   \\ Observatoire de Paris, LESIA, CNRS UMR 8109, F-92195 Meudon, France \\ \\
                   formerly at MPI for Astrophysics, Karl-Schwarzschild Str.\, 1, \\ 
          D-85741 Garching, Germany  \\ \vspace{2mm}   
           \\ J.\,Ballot \\ 
                   \\ Laboratoire d'Astrophysique de Toulouse-Tarbes, Universit\'e de            
                  Toulouse, CNRS, \\ 14 avenue Edouard Belin, F-31400 Toulouse, France \\ \\
                  formerly at MPI for Astrophysics, Karl-Schwarzschild Str.\, 1, \\ 
          D-85741 Garching, Germany  \\ \vspace{2mm}   
           \\ H.J.\,Muthsam \\ \\ Faculty of Mathematics, Nordbergstr.\, 15, \\
          A-1090 Vienna, Austria}
\date{April 30, 2009}          
\maketitle
\noindent
\begin{abstract}
We present results from 2D radiation-hydrodynamical simulations of fully compressible
convection for the surface layers of A-type stars with the {\sc ANTARES} code. 
Spectroscopic indicators for photospheric convective velocity fields show a maximum of
velocities near $T_{\rm eff} \sim 8000$~K. In that range the largest values are measured for 
the subgroup of Am stars. Thus far, no prognostic model, neither theoretical nor numerical, is 
able to exactly reproduce the line profiles of sharp line A and Am stars in that temperature
range. In general, the helium abundance of A stars is not known from observations. Hence, 
we have considered two extreme cases for our simulations: a solar helium abundance as an 
upper limit and zero helium abundance as a lower limit. The simulation for the helium free
case is found to differ from the case with solar helium abundance by larger velocities,
larger flow structures, and by a sign reversal of the flux of kinetic energy inside the hydrogen 
ionisation zone. Both simulations show extended shock fronts emerging from the optical
surface, as well as mixing far below the region of partial ionisation of hydrogen, and vertical 
oscillations emerging after initial perturbations have been damped. We discuss problems 
related to the rapid radiative cooling at the surface of A-type stars such as resolution 
and efficient relaxation. The present work is considered as a step towards a systematic
study of convection in A- to F-type stars, encouraged by the new data becoming
available for these objects from both asteroseismological missions and from
high resolution spectroscopy.
\end{abstract}

\section*{Introduction}

Research on A-type has benefitted enormously from the introduction of high resolution
spectrographs with $R \gtrsim 100,000$ and from advancements in
asteroseismology through telescope networks and satellite missions. They provide
a new way of probing our theoretical understanding of these objects and address
questions such as: how are convection and pulsation
linked to each other, in a physical parameter region no longer completely dominated
by the strong envelope convection found in cool stars such as our Sun? How does 
convective mixing in these stars affect diffusion processes? What physical mechanism
let chromospheres disappear in early A-type stars? How much does convection 
influence observed spectra and how does convection interfere with determining stellar 
parameters from spectroscopy?  This selection of physical questions demonstrates that
A stars provide a unique physical laboratory \citep{Landstreet04} underlying
the richness observed in their spectra \citep{Adelman04}. 

In the following we report on the first results from a project on numerical simulations
of A-type (and F-type) stars, motivated by the progress made by asteroseismological
missions such as CoRoT and MOST and challenged by results from high resolution
spectroscopy. The simulations ultimately aim at improving our understanding of
the dynamical processes of convection and its interaction with pulsation and,
through mixing, on diffusion processes that shape the appearance of A-type stars.
We first provide a background on how convection has observationally been identified
in A-type stars. We then summarise the theoretical explanations given for the peculiar
convective line broadening found in A stars and review previous attempts made to
model the surface convection zones in these stars. In the subsequent section we
discuss the setup of the numerical simulations we have performed for mid A-type
stars followed by a discussion of our results and our conclusions as well as
an outlook on further work.

\section*{Observational Background}

The existence of convection in the surface layers of A stars was first predicted
by \citet{Siedentopf33}. Initially, observational evidence for such convection zones came
from a non-zero microturbulence parameter $\xi$. Common values for $\xi$ were found
in the range of $2 \dots 4$~km~s$^{-1}$ for both normal A and Am-type main sequence
stars (see  \citealt{Landstreet07} for a review and \citealt{Adelman04} for examples
and further references). This readily exceeds the microturbulent velocities found for their
cool star counterparts (see \citealt{Gray88} for an extended discussion of $\xi$ in cool 
stars). To find direct evidence for convective velocity fields in the shapes of spectral
lines of A stars is much more difficult. The average projected rotational velocities of normal
A-type stars is found to be $> 120$~km~s$^{-1}$ and even for Am stars $v_{\rm e} \sin i$ 
values of less than $10 \dots 15$~km~s$^{-1}$ are the exception and not the 
rule (cf.\ the tables and references shown in \citealt{Adelman04} and also the recent
study of \citealt{Royer07}). Hence, first evidence that the bisectors of spectral lines in early
type (sharp line) stars do not resemble those observed for cool stars was found for
late F-type supergiants \citep{Gray86} and the early F-type bright giant Canopus
\citep{Dravins84,Dravins87}. Stars in this parameter range were shown to feature line profiles
with curved bisectors caused by extended depressions in the blue wing of spectral lines.
This results in an asymmetric line shape inverted in comparison with line profiles of cool 
supergiants (and those of our Sun and other cool main sequence stars). First evidence for this 
feature to be present in main sequence A stars was reported by \citet{GrayNagel89} for the 
case of HR\,178 (HD\,3883, an Am star). 

This was followed by a study of \citet{Landstreet98} who found two Am stars,
HR\,4750 (HD\,108642) and 32\,Aqr (HD\,209625), to show large asymmetries
and a depressed blue wing in profiles of strong spectral lines. Weak lines were found to
have much smaller profile asymmetries or even none at all. The same effects were
found to be slightly weaker in the case of the hotter A-type star HR\,3383 (HD\,72660).
A hot A-star with similar properties, HR\,6470\,A (HD\,157486\,A),  was identified by
\citet{Silaj05}. The current status of observational evidence of velocity fields caused by
surface convection in main sequence A stars is analysed in \citet{Landstreet09},
who added another five stars to this sample. The newly identified objects include
a mid A-type star with strong line profile asymmetries as previously found only for 
Am stars \citep{Gray89,Landstreet98}. Based on the collected data, which involved
also reobserving objects during the same night, in consecutive nights, in different
runs, and in different wavelength ranges, they concluded that the properties of the
line profiles of these stars are robust. Furthermore, they concluded that the only
explanation for the line profiles which is consistent with the observational data
is the presence of photospheric velocity fields not related to pulsation thus confirming
the suggestion by \citet{Landstreet98}. These velocity fields are present in both
normal A and in Am stars, though the latter contain the most extreme cases, in
agreement with their higher values of $\xi$. It is interesting to note that 
chromospheric activity indicators were identified for A-type stars with 
$T_{\rm eff} \lesssim 8200$~K \citep{Simon02} which roughly coincides with
the effective temperature of those stars for which the largest profile asymmetries
and overall convective line broadening have been identified \citep{Landstreet09}.

\section*{Theoretical Explanations and Previous Simulations}

The most straightforward explanation for the ``inverted'' or ``reversed'' line bisectors
has already been given by \citet{Gray86} for the case of supergiants who suggested
that this could be caused by hot columns of rising gas outshining the cold downflows. 
This was further developed by \citet{Gray89} and \citet{Dravins90} who noted that the 
hot upflows need to cover only a small fraction to reproduce such line profiles. This 
idea was also discussed for the case of main sequence A stars by \citet{Landstreet98}.

Could this be recovered from non-local models of convection or radiation-hydrodynamical
simulations? A first study of the coupled convection zones caused by ionization of
H~{\sc i} (and He~{\sc i}) as well of He~{\sc ii} was presented in \citet{Latour81}, who
used a modal expansion of an anelastic approximation to the hydrodynamical equations
to describe convective motions in a mid A-type star. They noted the two convection
zones to be coupled by overshooting. Due to the large He abundance (Y=0.354) assumed
in this work it is difficult to compare it with more recent calculations (the large value of $Y$
might partially explain why they found the convective flux to be larger in the He~{\sc ii}
ionisation zone, a property not recovered in later work). \citet{Xiong90} presented solutions 
obtained from his non-local model of convection for an F-type, an A-type, and two late B-type 
stars. Two separate convection zones were found coupled by overshooting for the mid
A-type star with a convective flux being larger for the photospheric convection zone. A solar
He abundance (Y=0.28) was assumed, as in the study of \citet{Kupka02}, who used the
fully non-local convection model of \citet{CD98} and \citet{Canuto01} to construct envelopes
of A stars for an evolutionary sequence of a 2.1~$M_{\odot}$ star and for an effective
temperature sequence at three different metallicities. Except for the hottest
A stars the two convection zones were found to be coupled by overshooting with the
photospheric zone dominating in mid and late A-type stars (in agreement with the 
numerical simulations by \citealt{Freytag95}, see below). They also investigated
photospheric overshooting and found a positive kinetic energy flux for the observable 
layers of mid and late A-type stars. Such a distribution agrees with the model
of \citet{Gray89,Dravins90,Landstreet98}, although no synthetic spectra were
computed at that point. 

The first radiation-hydrodynamical (RHD) simulations of a mid A-type star in two spatial
dimensions (2D) with $49 \times 42$ grid points were presented in \citet{Sofia84}
(radial resp.\ vertical coordinate listed first here and in the following). However, an
impenetrable upper boundary condition had to be placed at optical depths $\tau$ 
of 0.22 to $2/3$ in those simulations. This put the boundary layer right into the zone
of partial ionisation of H~{\sc i} with its accompanying density inversion and pushed 
vertical velocities to zero in that layer. Consequently, the lower convection zone
due to ionisation of He~{\sc ii} was found to dominate even at a solar He abundance
and no predictions of photospheric velocities could be made. The stably
stratified layers in-between those two zones were found to be fully mixed. In
\citet{Gigas89} a 2D RHD simulation for the parameters of Vega was presented.
The model had $30 \times 36$ points, but extended to much lower optical depths
of $\log \tau \sim -2.4$. No He ionisation was included in the equation of state 
and simulation time had to be limited to one hour of stellar time. Vertical oscillations
were found in the photosphere instead of a pattern of up- and downflows. Interestingly,
line profiles calculated using that simulation were found to be bent bluewards. However,
a comparison with \citet{Landstreet98} reveals that although the resulting bisector
looks similar to those found for strong lines in HD\, 72660, a strong asymmetry is
also found for weak lines contrary to observations, as corroborated by the 
analysis of \citet{Landstreet09}, who included another four stars in that $T_{\rm eff}$
range of 8900~K to 10000~K  in their analysis. Given the short simulation time and the
reported periods of 5 to 15~min the observed oscillations could also have been transient.
This is also implied by results of the first extended study of A stars by means of
2D RHD simulations by \citet{Freytag95} (see also \citealt{Freytag96}), who found 
oscillations to be damped even for a $T_{\rm eff} \sim 9000~K$ and a $\log(g)=3.9$, i.e.,
at the blue edge of the $\delta$~Sct instability strip, while oscillations were clearly 
excited in an even cooler model. Most of the simulation in \citet{Freytag95} were carried
out for a typical resolution of $65 \times 62$ grid points, with a few simulations doubling
horizontal resolution, and one case reaching $95 \times 182$ points for a short run. 
A solar chemical composition was assumed throughout and partial ionisation
was accounted for as well. The case of non-grey radiative transfer was considered
for one model and found to have little influence on the model structure and the
flow itself. The convective (enthalpy) flux was always found to be larger in the 
H~{\sc i} driven convection zone compared to the deeper one driven by  He~{\sc ii} 
(if included in the simulation box) for models with $T_{\rm eff} \lesssim 9000~K$ with
the two convection zones always connected by overshooting. A characteristic
property of the simulations were nevertheless rather high velocities in the lower
convection zone with vertical root mean square ($v_{\rm rms}$) maxima often
around 1~km~s$^{-1}$. These maxima are related to strong vortices found in the
lower convection zone. No synthetic spectra were computed from those simulations.

A surprising change to these developments came finally with the first RHD simulations
in 3D for two mid A-type main sequence stars by \citet{Freytag04} (with a $T_{\rm eff}$
of 8000~K and 8500~K for a $\log(g)$ of 4.4). The grey approximation and a solar chemical 
composition were assumed. 90 to 110 grid points were used along the vertical direction
and 180 points along the horizontal ones. Similar to the previous 2D RHD simulations
and solar granulation simulations they followed a box-in-a-star approach in which a limited
volume located near the surface of the star is considered for the calculations. Contrary
to most of the preceding 2D simulations the extent of the simulation boxes considered
by \citet{Freytag04} were chosen such as to contain at least several up- and downflow 
structures along any horizontal direction, which explains the larger number of grid points.
Nevertheless, their simulations revealed the usual granulation pattern, with some subtle 
differences. In the end this lead to line profiles very much looking like solar ones in 
contradiction to the observations by \citet{Landstreet98}. The calculated line profiles were 
further analysed by \citet{Steffen06}. In a follow-up work a third 3D RHD simulation (for 
a $T_{\rm eff}$ of 8000~K and a $\log(g)$ of 4.0) was presented by \citet{Kochukhov07}
using again the CO$^5$BOLD code as in \citet{Freytag04} and \citet{Steffen06}, but this 
time with a slightly higher resolution ($170 \times 220^2$). In addition, part of the run 
was repeated using non-grey radiative transfer (based on binned opacities). Due to the 
latter the match for weak lines of stars such as HD\,108642 (with actual stellar abundances
used for the spectrum synthesis) improved. The same held for the cores of strong lines, 
but the wings of strong lines remained clearly discrepant both in shape, width, and bisector curvature. Only one more 3D RHD simulation of a cool A-star ($T_{\rm eff} = 7300$~K, 
$\log(g) = 4.3$, solar metallicity, non-grey) with $82 \times 100^2$ grid points was presented
in more detail by \citet{Trampedach04}, though without computations of stellar spectra.
However, in that simulation the H~{\sc i}  and He~{\sc ii} zones are still connected to 
a single layer in a configuration described to be highly unstable. Further work in the
literature has dealt with simulating interacting convection zones assuming idealised 
microphysics, but these results cannot be directly compared to stellar observations
without invoking additional assumptions on the surface radiative cooling and the
effects of chemical composition.

The discrepancy found by \citet{Freytag04} is surprising, since RHD simulations using
the same code had been used to compute realistic solar spectra that match observed 
line profiles (see \citealt{Steffen07} and references therein). Obviously, the simulation
results are in contradiction with the model of hot and narrow, rapidly rising columns of
gas. The observations discussed by \citet{Landstreet09} indicate that chemical peculiarity 
\citep{Freytag04} can play only a limited role, as the line profile anomaly is found in 
both Am and normal A stars. This still cannot exclude He abundance to play a role 
\citep{Kupka05}, at least at the level of explaining the differences between stars 
showing the most asymmetric and broadened profiles (of Am-type) and their counterparts
among normal A stars. Effects of rotation might have to be considered, too (suggested
by Arlt in \citealt{Freytag04}, some consequences were analysed in \citealt{Kupka05}). 
This suggestion was also motivated by the frequent binary nature of the observed sharp
line stars. However, the line profiles in \citet{Landstreet98} and \citet{Landstreet09} 
do not resemble at all the flat-bottomed line profiles of rapid rotators seen pole-on
such as Vega \citep{Gulliver94,Hill04}. Since there is no indication from the data
on the sample of \citet{Landstreet09} that the binaries are very close or interacting,
nor that there is any dependence of the profiles on the rotation period other than the
ability of detecting them, rotation appears to be a less likely candidate to resolve
this discrepancy. Rather, numerical resolution \citep{Freytag04} or the entire scheme 
used for the radiative transfer (Landstreet, priv.\ comm.\ 2004) may play a much more
important role (see also \citealt{Kupka05}).

\section*{Numerical Simulations of A-type Stars with ANTARES}

To investigate the cause of these discrepancies between observation and numerical
simulations we have decided to perform RHD simulations of A-type stars with the 
ANTARES code (\citealt{Muthsam07}, \citealt{Muthsam09}). This work is part of a more
extended research project on convection in A- and F-type stars intended to study
convection-pulsation interaction and the properties of convection under physical conditions
that lead to very different efficiencies for the transport of heat and for the mixing of fluid, 
respectively. 

As a first step we decided to have a closer look at the effects of helium abundance and of
resolution, since both have not been investigated yet in sufficient detail. In addition, we also 
wanted to clarify the nature of the extremely restrictive time steps due to the short radiative 
cooling time scale $t_{\rm rad}$ \citep{Freytag04,Steffen07} (see Eq.~(\ref{eq.t-rad})). As 
ANTARES currently solves the fully compressible equations of hydrodynamics coupled to the 
stationary radiative transfer equation with a fully explicit time integration method, the ensuing 
time step constraints due to $t_{\rm rad}$ are severe and the resulting computational costs
very high (10 to 100 times higher than for a solar granulation simulation of comparable resolution
according to \citealt{Freytag04}). We have thus begun our study with RHD simulations in 2D
to single out the most promising cases for 3D simulations and possibly suggest improvements
to the time integration to reduce the computational costs of 3D simulations.

With ANTARES we solve the RHD equations for fully compressible flows,
\begin{equation}
\frac{\partial \rho}{\partial t} + \nabla \cdot \left[\rho {\bf u}\right] = 0,  \label{eq.cont}
\end{equation}
\begin{equation}
\frac{\partial \rho{\bf u}}{\partial t} + \nabla \cdot [\rho {\bf u}{\bf u} +p \underline{I}] = \rho{\bf g}  + \nabla \cdot \underline{\tau},  \label{eq.mom}
\end{equation}
\begin{equation}
\frac{\partial e}{\partial t} + \nabla \cdot [{\bf u}(e+p)] = \rho({\bf g}\cdot {\bf u})  + \nabla \cdot ({\bf u} \cdot \underline{\tau}) + Q_{\rm{rad}},  \label{eq.energy}
\end{equation}
which are the continuity, Navier-Stokes, and total energy equation describing the conservation 
of mass density $\rho$, momentum density $\rho {\bf u}$, and total energy density $e$.
The later is the sum of internal ($e_{\rm int}$) and kinetic energy density, respectively, and 
${\bf u}$ is the flow velocity. An equation of state (EOS) relates $e_{\rm int}$ and $\rho$ to the
temperature $T$ and the gas pressure $p$ (or rather, the sum of gas pressure and isotropic
radiation pressure $p_{\rm rad}$, the meaning of $p$ as used in the following). The EOS is taken 
from the most up-to-date OPAL tables \citep{Rogers96} and used also to compute all other required thermodynamic derivatives. We note that $\underline{I}$ is simply the identity matrix, 
$\underline{\tau}$ is the viscous stress tensor, and $t$ and ${\bf x}$ denote time and space. 
The simulations discussed below are box-in-a-star simulations for a limited volume
ranging from the upper photosphere to the stably stratified layers around $T \sim 60,000$~K 
to $70,000$~K, thus containing the surface convection zones inside the box. 
Since this region is small with respect to the stellar radius, a Cartesian (box) geometry
is taken for the simulation box and the gravity ${\bf g}$ is a constant, downwards pointing
vector equalling the surface gravity $g$. Horizontal boundary conditions are hence periodic
and the geometry is plane parallel. The horizontal domain is chosen wide enough to fit 
several up- and downflow structures into the box at photospheric depth. The vertical
boundary conditions are assumed to be impenetrable and stress-free (free-slip) with 
a constant energy flux assumed at the bottom of the simulation zone (and directly related
to the $T_{\rm eff}$ defined for the simulation). The lower boundary is placed sufficiently far 
below the convection zones to have only limited influence (at most through reflecting waves). 
The upper boundary condition is placed at $\log \tau_{\rm ross} \sim -5$ for similar reasons.
The appropriate domain sizes are estimated from scaling the linear dimensions of solar 
granulation simulations by the pressure scale height $l = H_p = P / (\rho g) \sim T / (\mu g)$, 
which when comparing to the Sun implies
\begin{equation}  \label{eq.scaling}
l / l_{\odot} = T_{\rm eff} \mu_{\odot} g_{\odot} / (T_{\rm eff,\odot} \mu g)
\end{equation}
for a star characterised by 
$(T_{\rm eff},\mu,g)$, where $\mu$ is the mean molecular weight. In optically thick layers
the radiative (heating and) cooling rate $Q_{\rm rad} = \nabla \cdot F_{\rm rad}$ can be computed from the diffusion approximation $F_{\rm rad} = -K_{\rm rad}\nabla T$, where $K_{\rm rad}$
is the radiative conductivity obtained from equation of state quantities and the Rosseland
opacity $\kappa_{\rm ross}$. For the latter, we use tabulated OPAL opacities \citep{Iglesias96} 
extended at low temperatures with tables by \citet{Ferguson05}, both computed for the
solar mixture of \citet{Grevesse93}. For the photospheric
layers a radiative transfer equation must be solved for accurate computation of $Q_{\rm rad}$.
This computation is extended into deeper layers to allow a smooth transition to the diffusion
approximation. Details on this procedure as used in ANTARES are given in 
\citet{Obertscheider07} and in \citet{Muthsam09}. We note that even in the grey approximation
the results of this approach are different from the plain diffusion approximation. In ANTARES
the stationary limit of the radiative transfer (intensity) equation is assumed (as in
\citealt{Freytag95}, \citealt{Freytag04}, \citealt{Kochukhov07}) and the resulting $Q_{\rm rad}$
is computed from an integration of the intensity obtained for a finite set of rays from
this procedure. The grey case differs from the non-grey one by just integrating along
each ray for a single frequency band rather than for multiple ones, but contrary to the
diffusion approximation this accounts for inhomogeneity of cooling and heating 
in a convective stellar photosphere. We also point out here that the assumption
of stationarity (instantaneous radiative transfer) leaves $\rho$, $\rho{\bf u}$, and $e$
as the dynamical variables of the system which have to be followed by the time
integration during the simulation.

To perform simulations which are close to an observational case we have chosen 
$T_{\rm eff} = 8200$~K and a $\log(g) = 4.0$ as basic physical parameters, as they 
are representative of HD\,108642 (\citealt{Kupka04}; note that \citealt{Landstreet98} and
\citealt{Landstreet09} suggested a $T_{\rm eff}$ 100~K lower and a $\log(g)$ 0.1~dex
higher than those values, but such differences are well within the observational
uncertainties). Since diffusion has been suggested to substantially change the helium
content of the upper envelope during the main sequence life time of A-type stars 
\citep{Vauclair74, Richard01,Michaud04}, we consider two extreme cases of helium 
abundance: a solar one and no helium at all. A solar metallicity mass fraction of $Z=0.02$ 
was taken in both cases (this value is a compromise, because the underabundance
of light elements relative to their solar values lowers the absolute value of $Z$ while the
opacity resulting from the different chemical composition contributing to $Z$ is increased
in the photosphere of Am-type stars; see \citealt{Kupka04} for further details on the 
abundances of HD\,108642). The hydrogen mass fraction $X$ is scaled up from 0.7 to 0.98
in the case where helium is absent ($Y=0$). 

For the simulations presented here we have chosen a coarse grid with a single refinement
zone embedded inside the domain covered by the coarse grid. Since the simulations are
in 2D, the horizontal boundary conditions of the coarse grid are periodic and because the fine
grid is used to improve resolution near the surface, it has the same horizontal extent as the 
coarse one. Hence, it has periodic boundary conditions as well. Vertically, the fine grid is
located in the interior of the coarse grid which provides the vertical boundary conditions for
the fine grid. The coarse grid itself has the closed vertical boundary conditions already
described above. For the case with solar helium abundance, the coarse grid has
$150 \times 200$ points which span a domain of $32.2 \times 40.3$~Mm$^2$ and yield a 
resolution of $216 \times 201$~km$^2$. The fine grid has $205 \times 200$ points and is
located vertically between about 1.1~Mm and 12.1~Mm as measured from the top of the 
simulation box. This provides a resolution of $54 \times 201$~km$^2$, i.e., a vertical
refinement by a factor of 4. For the plots shown below the vertical axis is shifted by
$-3.8$~Mm to have the downwards pointing vertical axis be labelled with zero where the 
horizontally and temporally averaged temperature equals $T_{\rm eff}$. For the simulation
without helium, a coarse grid of $130 \times 210$ points is used spanning a domain of 
$26.1 \times 41.7$~Mm$^2$ and thus providing a mesh size of $202 \times 199$~km$^2$.
The fine grid is located vertically between about $1.8$~Mm and $13.1$~Mm as measured
from the top and has $225 \times 210$ grid points. Hence, the resolution on the fine grid 
is $51 \times 199$~km$^2$, which again implies a vertical refinement by a factor of 4.
For this simulation the vertical axis is shifted in the plots shown below by $-4.6$~Mm.
As a result, in plots showing both simulations the vertical axis value of zero indicates
the average location where $T=T_{\rm eff}$, i.e., the optical surface, for both cases.
For the radiative transfer, we have used 12 rays and a single bin for Rosseland mean
opacities, i.e., the grey approximation.

Apart from setting up the simulations and some test runs which were done on single
and two CPU core configurations most of the simulation runs to evolve both the
solar helium and the helium-free case were done on 8~CPU cores using the MPI
capabilities of ANTARES. The code was slightly modified to allow grid refinement
zones to cover the entire horizontal extent of the model, as the fine grid was previously
used to have higher resolution only in a specific domain of interest (cf.\ \citealt{Muthsam07}).
These enhanced possibilities to use fine grids are now a standard feature of the code
\citep{Muthsam09}. Another improvement added to the code and motivated by simulations of
A-type stars was its ability to work with arbitrary chemical compositions (and thus combinations
of $X$,$Y$, and $Z$), since the opacity and equation of state tables used have some 
restrictions on the combinations of these parameters. The improved microphysics interface 
is now also a standard module of the code. After initializing the simulations, all production
runs were performed on the POWER5 p575 of the RZG in Garching. The initial conditions for
the simulations were taken from 1D models for HD\,108642 calculated with the LLmodels
code \citep{Shulyak04} using the convection model by \citet{CM91} and opacity distribution
function tables by \citet{Kur93a} and \citet{Kur93b}. For the 1D models $\xi = 4$~km~s$^{-1}$
was assumed \citep{Landstreet98,Kupka04} and $T_{\rm eff}$ and $\log(g)$ were the same as 
in the simulations (values taken from \citealt{Kupka04}). The flow in the simulations is started
by a lack of perfect hydrostatic equilibrium of the stratification introduced by the differences
in the equation of state and opacities used, the change in grid spacing between the 1D model
and the simulation, and the different numerical methods used in both codes. In addition,
we add a small perturbation in the mass density field by creating a random distribution for the 
density perturbations in Fourier space. The values of this smooth noise function are added at 
each point to the average for the horizontal layer as obtained from the 1D model. The noise is 
equally distributed over all modes (wave numbers), but its magnitude is scaled as a function of 
depth to produce the largest density fluctuations where the 1D convection model predicts the 
largest temperature fluctuations. This avoids the introduction of long-lived perturbations in the 
radiative region near the bottom of the simulation box, which would cause unnecessarily long 
relaxation times (we return to this topic further below).

\section*{Results}

In the following we discuss the main results of our 2D RHD simulations for both the case
with solar helium abundance and for the case with zero helium abundance. 
Some results from experimental runs with different initial conditions and resolution
but otherwise identical parameters are included as well.

\subsection*{Relaxation and oscillations}

\begin{figure}[!th]
\begin{center}
\epsfig{file=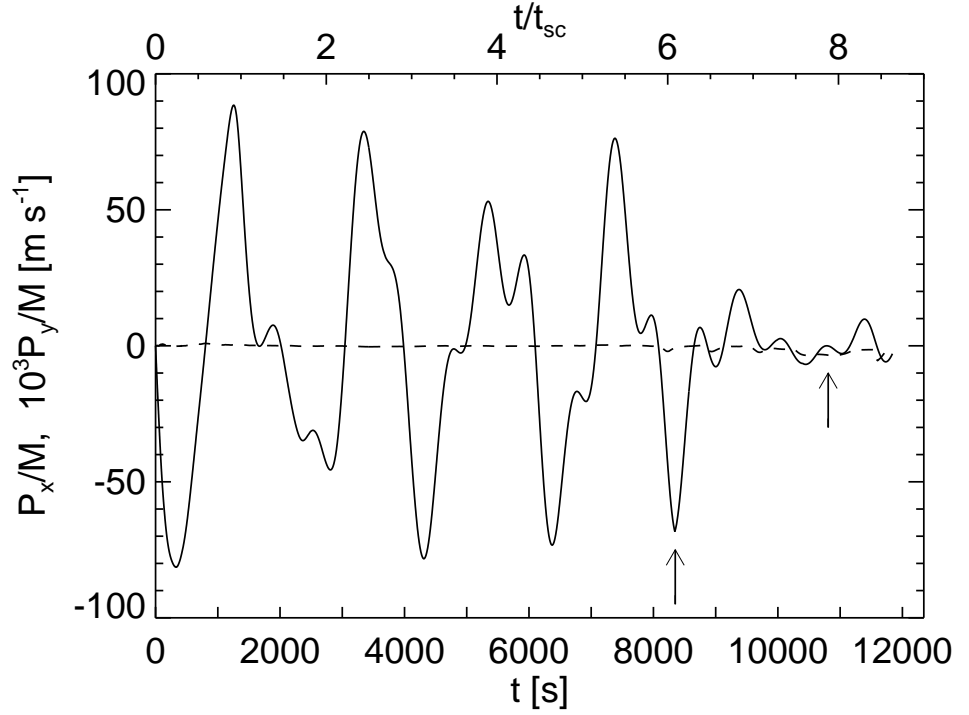,clip=,angle=0,width=\linewidth}
\caption{Vertical (solid line) and 
    horizontal (dashes) velocities of the centre of
    gravity for the simulation with solar helium abundance as a function of time. 
    The horizontal component is magnified by a factor of $10^3$. On the top x-axis,
    time is normalised by the sound-crossing time. Arrows indicate the beginning 
    and the end of the artificial damping phase.}
\label{Fig_cg_vel_He}
\end{center}
\end{figure}

Both simulations were run for several sound-crossing times until convective motions had 
sufficiently developed, which is indicated by a growth in vorticity and in total kinetic energy. 
For the simulation with solar He abundance notable growth
of convective motions occurred after about three sound-crossing times 
(time for a sound wave to cross the entire simulation box vertically, 
here $t_{\rm sc} = 1371.6$~s).
After $t/t_{\rm sc} \gtrsim 5$ convective motions dominated the contributions
to kinetic energy and experienced a phase of exponential growth. Figure~\ref{Fig_cg_vel_He}
shows the velocity of the centre of gravity of the entire simulation volume as 
a function of time. The oscillations triggered by the stratification being out of perfect
hydrostatic equilibrium are prominent and are almost purely vertical since the horizontal component is totally negligible. There is only weak intrinsic damping. 
We attribute this to the quality of the advection scheme implemented
in ANTARES. We removed these oscillations by introducing a damping term
at $t/t_{\rm sc}  \sim 6.1$ for the vertical velocity component, as is routinely done so during
relaxation of solar granulation simulations \citep{Trampedach97}. Due to the short time
scale chosen the oscillations are rapidly damped which allows us to turn off this artificial 
damping after $t/t_{\rm sc} \sim 7.9$. At this point the vertical oscillations contribute
$\lesssim 3\%$ to the total kinetic energy of the flow and shortly afterwards the total kinetic 
energy starts dropping for the first time after its rapid phase of growth and at least the upper 
convection zone itself is essentially relaxed. The snapshots in Figs.~\ref{Fig_tf_He},\; 
\ref{Fig_choc_a}, and~\ref{Fig_choc_b} have hence been taken at $t/t_{\rm sc} \gtrsim 8$.
The equivalent holds for the averages in Fig.~\ref{Fig_gradients_He} and~\ref{Fig_fluxes_He}.

\begin{figure}[!th]
\begin{center}
\epsfig{file=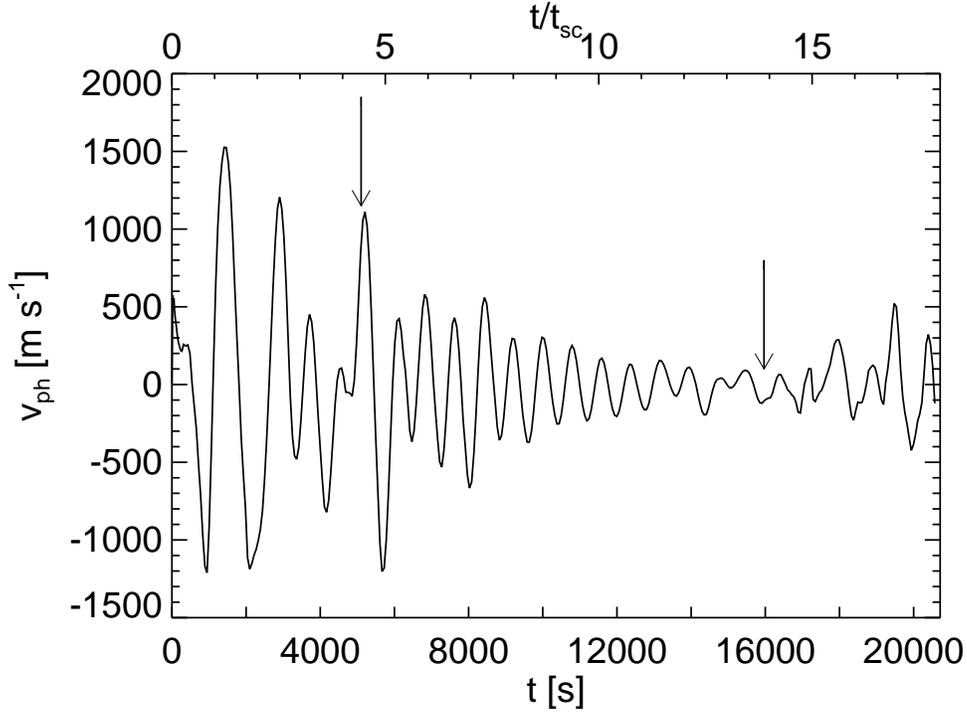,clip=,angle=0,width=\linewidth}
\caption{Temporal evolution of 
    photospheric velocity for the simulation with zero helium abundance. On the top x-axis
    time is normalised by the sound-crossing time. Arrows indicate the beginning 
    and the end of the artificial damping phase.}
\label{Fig_ph_vel_noHe}
\end{center}
\end{figure}

An alternative way of visualizing vertical oscillations is to look at the motion of the
photosphere and define the velocity $v_{\rm ph}$ as the horizontally averaged vertical 
motion of the layer for which $T=T_{\rm eff}$ (the photospheric radius) at a given point in
time. For both simulations this velocity is about an order of magnitude larger than that of
the centre of gravity. Figure~\ref{Fig_ph_vel_noHe} shows the time development of
$v_{\rm ph}$ for the simulation with zero helium abundance. Intrinsic damping is just slightly
more efficient than in the previous case.  Artificial damping of vertical motions is applied
from $t/t_{\rm sc} \sim 4.4$ to $t/t_{\rm sc} \sim 13.9$ onwards (here, $t_{\rm sc} = 1150.9$~s). 
As we have used a smaller damping rate for this second simulation the damping phase is
longer and smoother than for the previous case. The development of convection also takes 
longer: strong convective motions set in only after $t/t_{\rm sc} \sim 7$ and they dominate
over the vertical oscillations in terms of kinetic energy after $t/t_{\rm sc} \sim 9$.
While the convective motions increase in strength, the contributions by oscillations  
drop until they reach a level of $\lesssim 3\%$ at $t/t_{\rm sc} \sim 14$. Once the 
flow can evolve undamped, some oscillatory motions seem to rise again. This is more
obvious for $v_{\rm ph}$ than for the centre of gravity, despite $v_{\rm ph}$ is also much
more influenced by events taking place at the surface such as the formation of shock
fronts, which contribute to the non-sinusoidal shape of the function visible in 
Fig.~\ref{Fig_ph_vel_noHe}. With kinetic energy increasing no longer exponentially
the simulation slowly reaches equilibrium at $t/t_{\rm sc} \sim 18$. 
The snapshot in Fig.~\ref{Fig_tf_noHe} and the averages in Fig.~\ref{Fig_fluxes_He} 
have been computed for $t/t_{\rm sc} \gtrsim 17$. 

We notice that a proper choice of the initial perturbations is
more important for simulations of surface convection in A stars than for solar granulation
simulations. An earlier model for the case of solar helium abundance, which had been
perturbed with an additional, sinusoidal velocity field instead of a random perturbation
in density, showed strong resiliency to ``forget'' this pattern in the zone of He~{\sc ii}
ionisation even after several sound-crossing times. That simulation run was hence
discarded. This behaviour is quite different from the solar case where the same kind
of perturbation is rapidly removed by the convective flow, for instance, in the simulations 
of \citet{Obertscheider07}.

\subsection*{Structure of the convection zone}

\begin{figure}[!th]
\begin{center}
\epsfig{file=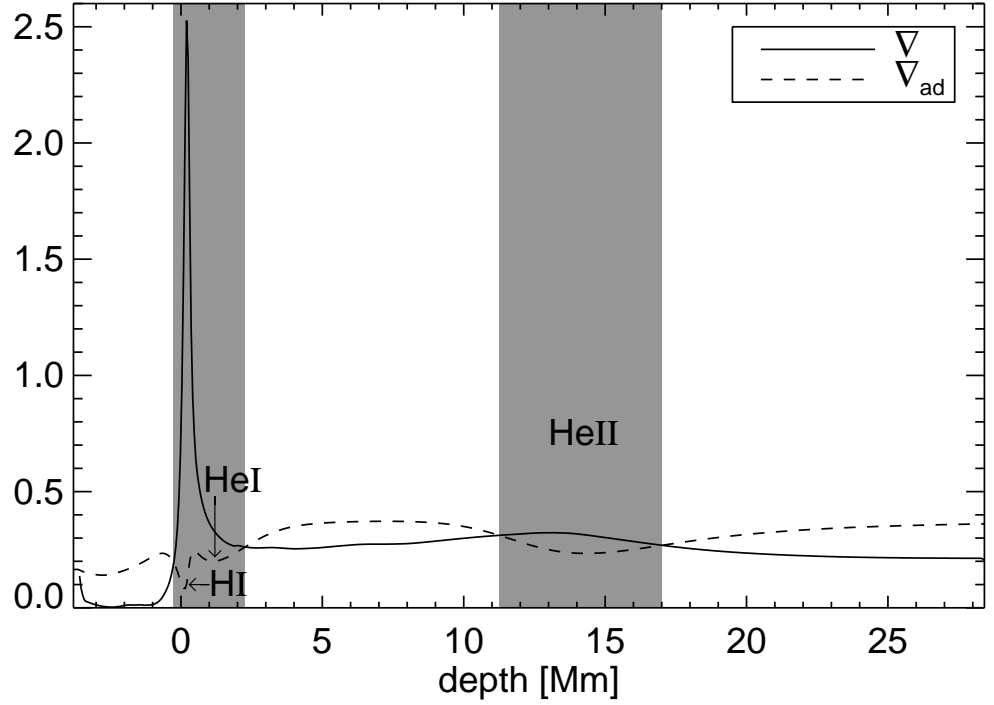,clip=,angle=0,width=\linewidth}
\caption{Temporal and horizontal average of actual 
    ($\nabla$) and adiabatic ($\nabla_{\rm ad}$) temperature gradient for the simulation
    with solar helium abundance. The regions of partial ionisation of H~{\sc i}, He~{\sc i},
   and He~{\sc ii} are indicated. Layers unstable according to the Schwarzschild
   criterion are indicated in grey.}
\label{Fig_gradients_He}
\end{center}
\end{figure}

In Figure~\ref{Fig_gradients_He} we show the dimensionless temperature gradient
$\nabla$ and the dimensionless adiabatic gradient $\nabla_{\rm ad}$ for the simulation
with solar helium abundance. The superadiabatic peak in the zone of partial hydrogen
ionisation is about 3.5 times steeper than in a simulation of solar granulation 
(cf.\ \citealt{Rosenthal99} for the latter). The upper convection zone is about
2.52~Mm deep as obtained from the definition of the Schwarzschild criterion 
applied to the horizontally and time averaged gradients. The upper zone is mostly
driven by the partial ionisation of H~{\sc i}. It is extended due to the partial ionisation 
of He~{\sc i} which takes place close enough to unite both regions into a single
convection zone. With 5.7~Mm the second convection zone, which is caused by
partial ionisation of He~{\sc ii}, is much larger in vertical extent. This is mostly due
to the increase of the local pressure scale height. The resulting mean structure is
very similar to that one found in one dimensional models of stellar atmospheres
and stellar envelopes for that part of the HR diagram (cf.\ \citealt{Vauclair74},
\citealt{Latour81}, \citealt{Landstreet98}), because the high radiative losses of
convection keep the stratification of the entire envelope in these stars close to that
one of a purely radiative model (this has also been found by \citealt{Freytag95}, 
\citealt{Freytag96}, \citealt{Kupka02}, \citealt{Freytag04}). 

\begin{figure}[!t]
\begin{center}
\epsfig{file=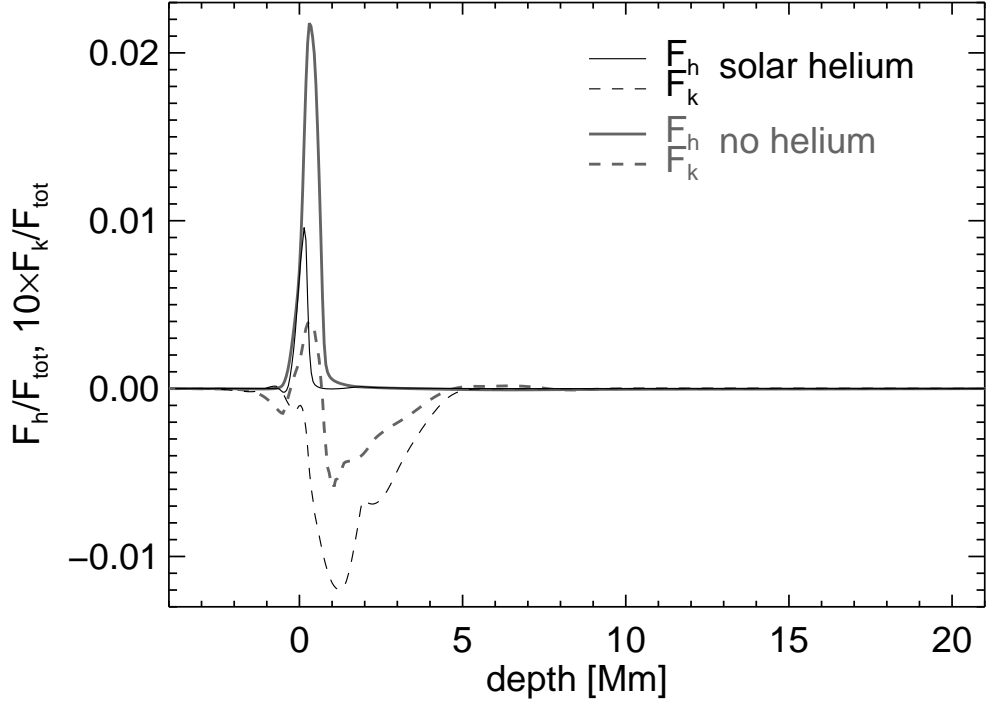,clip=,angle=0,width=\linewidth}
\caption{Convective (enthalpy) and
    kinetic energy flux in units of the total flux, shown as solid and dashed line, respectively,
    for the simulation with solar helium abundance (black thin) and without helium
    (grey thick). The kinetic energy flux is multiplied by a factor of 10 for better visibility.}
\label{Fig_fluxes_He}
\end{center}
\end{figure}

By comparison, the simulation with zero helium abundance has just a single convection
zone due to the partial ionisation of H~{\sc i}. Its extent into the photosphere up from where
on average $T = T_{\rm eff}$ is marginally smaller than in the previous case (0.23~Mm
instead of 0.25~Mm). With a total extent of 1.56~Mm when following the same definition as
used above, it is more shallow than its counterpart containing helium, in spite of a larger
local scale height due to a lower $\mu$, Eq.~(\ref{eq.scaling}). This is a direct result of the
lack of He~{\sc i} ionisation extending the zone deeper inside (note the behaviour of
$\nabla_{\rm ad}$ in Fig.~\ref{Fig_gradients_He}). However, with a maximum of 1.6
the gradient $\nabla$ itself is flatter and the resulting superadiabatic gradient is only 
about 2.3 times steeper than in solar granulation simulations. 

\begin{figure}[!t]
\begin{center}
\epsfig{file=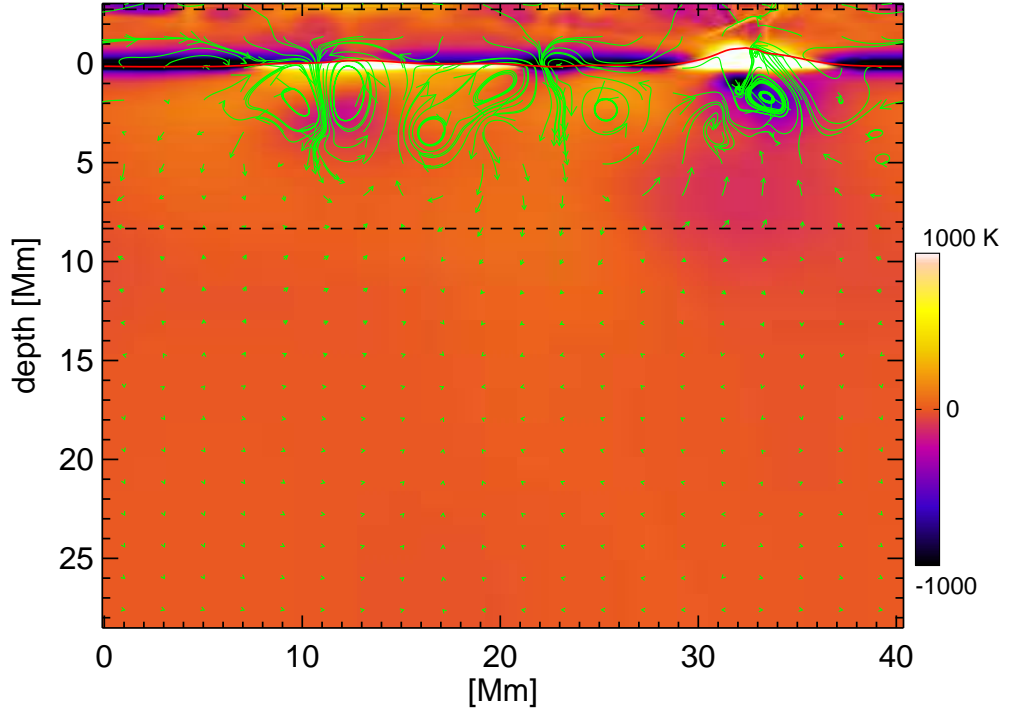,clip=,angle=0,width=\linewidth}
\caption{Snapshot of temperature fluctuations 
   ($[T']=K$) relative to their horizontal mean for the simulation with solar helium
   abundance. Extreme values have been chopped for easier visualization. The two
   black dashed lines indicate the region where grid refinement has been applied. 
   The $\tau_{\rm ross}=1$ `isosurface' is denoted by a solid line. Streamlines indicate the 
   direction and magnitude of the local flow field (the mean vertical velocity has been
   subtracted to improve visibility of the weakest vortices).}
\label{Fig_tf_He}
\end{center}
\end{figure}

Differences are also visible in the flux distribution. Figure~\ref{Fig_fluxes_He}
displays the enthalpy (convective) flux $F_{\rm h}$ and the flux of kinetic energy
$F_{\rm k}$ for both simulations in units of the total (input) flux after subtracting the 
contribution due to the mean flow (i.e., due to the remaining vertical oscillations).
That contribution is negligible for $F_{\rm k}$, but can contribute a relative flux of
up to 0.5\% to $F_{\rm h}$, if the averaging time is short, i.e., $t < t_{\rm sc}$.
The higher enthalpy flux found in the simulation with no helium is hardly significant
since it could be the signature either of a more relaxed simulation or of an effectively
stronger convection due to a lower mean molecular weight. The same could hold for 
the higher velocities found for the helium free case, though we consider them 
more likely to be caused indeed by a lower $\mu$. Nevertheless, and interestingly 
enough, we note that in the simulation with no helium $F_{\rm k}$ has the opposite
sign for the upper half of the convection zone. This feature remains robust
even when averaging for $t/t_{\rm sc} \sim 3$ and also when averaging over one third 
of that time scale for different subintervals. This inversion in the sign of 
$F_{\rm k}$, which does not appear in simulations of solar convection, occurs in our 
simulation with the most shallow convective zone. In this case with a really thin
convective layer heating from below and cooling from above act at the same time
within a pressure scale height, while in a thicker zone, such as in the model with
solar helium abundance, heating and cooling are more disconnected
(cf.\ \citealt{Moeng90} for a discussion of heating from below and cooling from
above in a meteorological context). We also note 
that at the chosen scale the fluxes in the He~{\sc ii} ionisation zone are not visible 
(cf.\ \citealt{Steffen06} and \citealt{Steffen07}). These small fluxes require a more careful 
inspection of the simulation. 

Indeed, if we look at Fig.~\ref{Fig_tf_He} we see that the strongest vortices created
by the flow are located in the upper convection zone or just slightly underneath it
and thus are close to the region of the strongest convective driving. The large
downdraft at a horizontal coordinate of 23~Mm manages to penetrate just a little bit
into the lower convection zone, but at that time no strong vortices have developed 
in that region. We recall our previous observation that if vortices are seeded there
as an initial condition, for instance by a sinusoidal perturbation, they remain there
for a long time. This is both due to the weak convective driving (which cannot rapidly 
destroy a vortex or push it into a stably stratified layer) and due to the general longevity
of vortex patches in 2D (cf.\ \citealt{Muthsam07}). Temperature fluctuations are found 
largest at the optical surface (the extrema have been truncated in the plot for a better
contrast and can be four to five times larger while root mean square fluctuations are 
twice as large). The optical surface is corrugated mostly around rapidly 
evolving upflow regions. 

\begin{figure}[!t]
\begin{center}
\epsfig{file=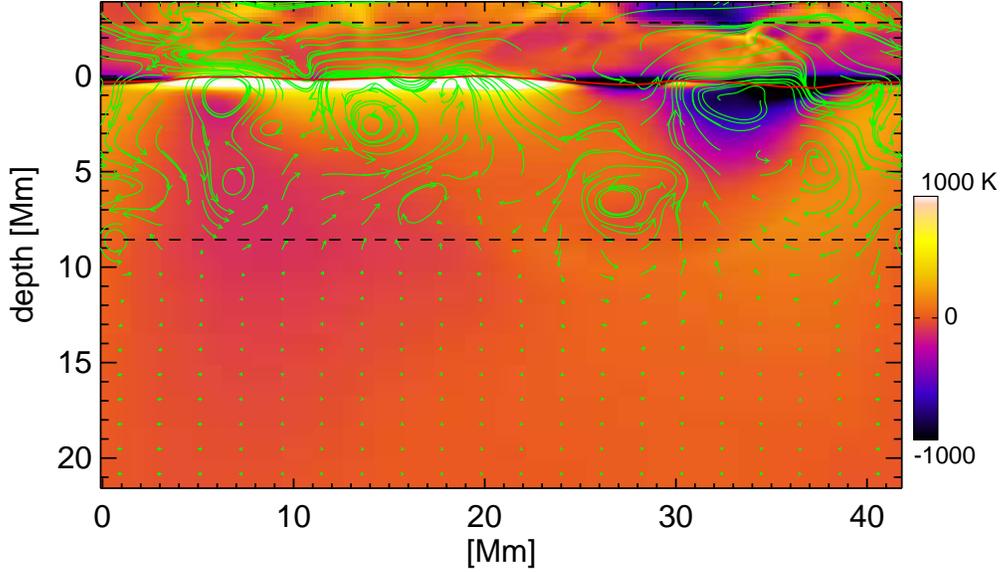,clip=,angle=0,width=\linewidth}
\caption{Snapshot of temperature fluctuations 
  ($[T']=K$) relative to their horizontal mean for the simulation without helium.
  Scaling of $T'$, domain boundaries of the grid refinement zone, the location of
  the optical surface defined at $\tau_{\rm ross}=1$, and the local velocity field 
  (with mean vertical velocity subtracted) are indicated as in Fig.~\ref{Fig_tf_He}.}
\label{Fig_tf_noHe}
\end{center}
\end{figure}

In Figure~\ref{Fig_tf_noHe} we see quite a similar overall pattern for the simulation
without helium. The horizontal scales are somewhat larger as expected from 
Eq.~(\ref{eq.scaling}). Vortex patches are also strongest inside the convection zone
itself, but a larger number of them is found further below than in Fig.~\ref{Fig_tf_He}. We
think that this is a result of the longer time evolution of this simulation. Eventually,
also in Fig.~\ref{Fig_tf_He} downflows from above will form new vortex patches
in the lower convection zone or patches located in the stably stratified layer in
between will reach it. Thus, if we run the simulation with solar helium
abundance significantly longer, we may eventually see an increase in $F_{\rm h}$.
However, this may not be more realistic, as the vortex patches in 2D just 
happen to assemble near the bottom of a simulation of a strongly stratified medium
(cf.\ \citealt{Muthsam07}). Since 3D simulations started from 2D ones may inherit 
these properties for quite some (simulation) time, we think that future long-term simulations
are the only way to accurately compute the enthalpy and kinetic energy fluxes in the lower
convection zone of models for A stars with a helium content large enough to drive 
that zone.

\subsection*{Shock fronts and their development}

\begin{figure}[!th]
\begin{center}
\epsfig{file=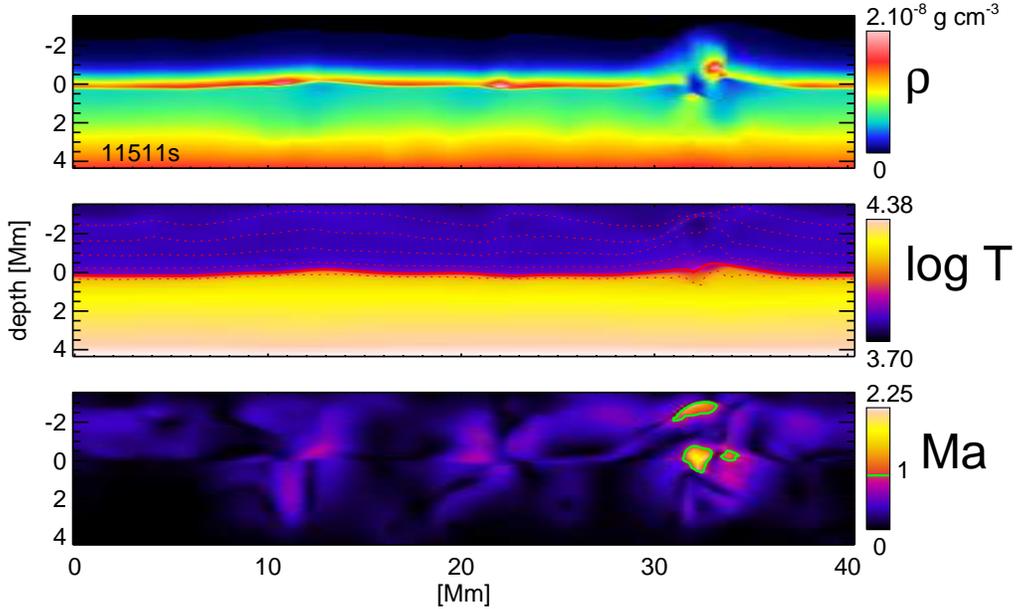,clip=,angle=0,width=\linewidth}
\caption{Mass density $\rho$, logarithm of temperature
     $T$ ($[T]=K$), and local Mach number in the simulation with solar helium abundance
     prior to the formation of a large shock front. The isoline for which $\tau_{\rm ross}=1$
     is denoted in the middle panel by a thick solid line. Isolines for $\log \tau_{\rm ross} =
     -4, -3, -2, -1, +1$ are denoted as dotted lines. Note the density inversion near the
     optical surface in the top panel. Regions of supersonic flow are surrounded by a bright
     solid isoline where ${\rm Ma}=1$.}
\label{Fig_choc_a}
\end{center}
\end{figure}

As we can conclude from the small enthalpy fluxes found for both simulations shown in 
Fig.~\ref{Fig_fluxes_He} convection is rather inefficient in transporting heat in mid A-type stars,
a result also reported by \citet{Freytag95}, \citet{Freytag96}, \citet{Kupka02}, \citet{Freytag04},
\citet{Steffen06}, and \citet{Steffen07}. Consequently, the mean temperature gradient stays
close to the radiative one and hence, contrary to RHD simulations of solar granulation,
a density inversion appears in the layer where H~{\sc i} is partially ionised \citep{Freytag95}.
This is a consequence of the lower mean molecular weight which in turn is caused by the
doubling of the number of particles contributed by ionised hydrogen relative to neutral one.
In the Sun efficient convection and thus a much flatter $\nabla$ is counterbalancing
this effect of partial ionisation (cf.\ the simulations of M.~Steffen and F.J.~Robinson compared
in \citealt{Kupka09}; the model by \citealt{CM91} is an exception, since it predicts inefficient 
convection for the photospheric layers of the Sun which are located on top of a rapid transition
to efficient, adiabatic convection dominating the bulk part of the convective envelope).  
Figure~\ref{Fig_choc_a} shows a snapshot for our simulation with solar helium abundance.
The density inversion is prominent and the ensuing Rayleigh-Taylor instability enhances
the convective instability caused by high opacity and a low $\nabla_{\rm ad}$
(see Fig.~\ref{Fig_gradients_He} for the latter). 

\begin{figure}[!t]
\begin{center}
\epsfig{file=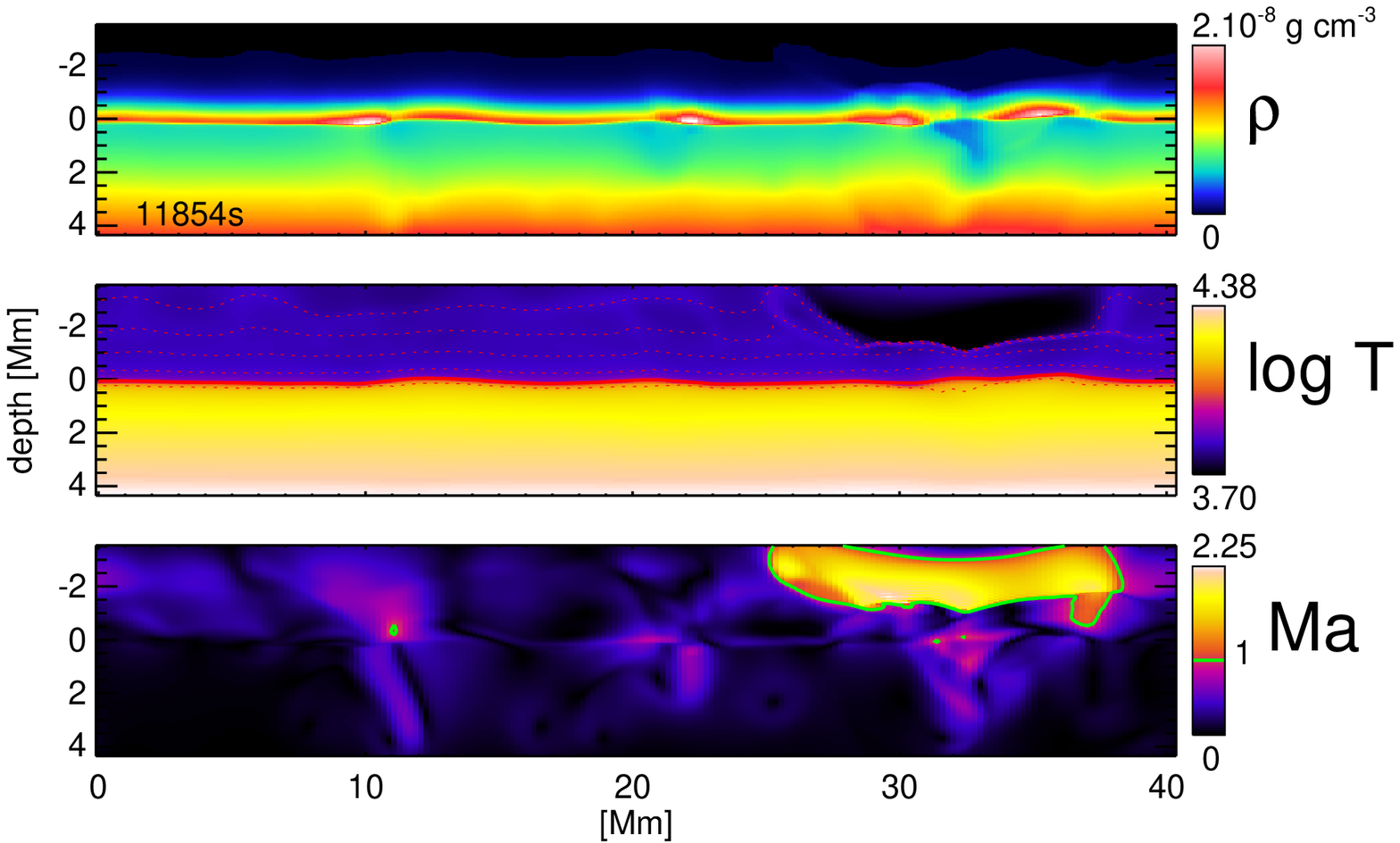,clip=,angle=0,width=\linewidth}
\caption{Mass density $\rho$, logarithm of temperature
     $T$ ($[T]=K$), and local Mach number in the simulation with solar helium abundance
     343~s ($0.25\,t_{\rm sc}$) after the time step displayed in Fig.~\ref{Fig_choc_a}. Note
     the large shock front in the photosphere spreading more than $1/3$ of this layer within
     the simulation box. It is characterised by large jump in temperature and a relative increase
     in the optical depth by more than an a factor of 10. Isolines have the same meaning as
     in Fig.~\ref{Fig_choc_a}.}
\label{Fig_choc_b}
\end{center}
\end{figure}

Note that the inversion layer is interrupted by a complex feature at a horizontal coordinate 
of $\sim 33$~Mm. In this region the optical surface is more corrugated (solid line in the 
temperature plot in the middle panel) and supersonic velocities are found there, too (bottom 
panel). This feature is the result of two counterrotating vortices, which form at some time 
$t/t_{\rm sc} \gtrsim 7.5$. At $t/t_{\rm sc} \sim 8$ they have created an extended region
just underneath the photosphere with a temperature lower than its horizontal average. The low
temperature fluid is supplied by the strong downdraft between the vortex pair which reaches 
supersonic velocities in its inflow area within the photosphere. At $t/t_{\rm sc} \sim 8.2$ hot
gas from the neighbouring upflows has mostly covered the cold downflow region leaving
a dense spot of cold material in its centre. As a result supersonic flow appears both in the
two photospheric inflow areas and in the downdraft itself. From now on the structure begins
to collapse emitting a series of shock fronts. At $t/t_{\rm sc} \sim 8.4$, which is shown in
Fig.~\ref{Fig_choc_a}, there are no more supersonic inflow regions. Rather, supersonic
velocities are found in a small front in the upper photosphere (bottom panel), visible also in
$T$ (band structure trailed by material of lower $T$ underneath) and in $T'$ in 
Fig.~\ref{Fig_tf_He}, which shows the state of the simulation 68~s before that one
illustrated in Fig.~\ref{Fig_choc_a}. The streamlines in Fig.~\ref{Fig_tf_He} also 
demonstrate that the structure is no longer dominated by a pair of roughly equally strong
vortices. Supersonic flow is also found just underneath the patch of dense gas extending
into the photosphere (Fig.~\ref{Fig_choc_a}).

Once this gas patch has bumped onto the layers just below it, a large shock front is created 
which horizontally extends over one third of the simulation box and also spreads over almost
one third of the photosphere contained in the simulation volume. This can be seen in
Fig.~\ref{Fig_choc_b}, at $\sim 0.25\,t_{\rm sc}$ after the state shown in Fig.~\ref{Fig_choc_a}.
The snapshot shows the moment when the remainders of the previous front, which is shown
still moving upwards in Fig.~\ref{Fig_choc_a} and is reflected by the upper boundary condition
shortly afterwards, are absorbed by the new front. This explains the extended high Ma number 
region preceding the new front (we recall here that a shock front is a (near) discontinuity in the 
hydrodynamic variables, which does not necessarily have to move at supersonic velocities). As 
the upwards moving material has much higher density, it immediately absorbs the downwards moving one with very little changes to its physical state. Note that the shock 
front is optically thin, although with a steep gradient in optical depth 
(isolines of $\log \tau_{\rm ross} = -3$ and $-4$ nearly merging with each other). At the
same time the anomaly in the density stratification has nearly disappeared. Comparing
with Fig.~\ref{Fig_choc_a} we can see that the density in this region has dropped 
dramatically (in fact to average horizontal values) leaving a low temperature region which
is reheated by the upwards moving shock front.
We note that similar events also appear in the simulation run without helium and smaller
events appear more frequently than the extreme case illustrated by Fig.~\ref{Fig_choc_a} 
and Fig.~\ref{Fig_choc_b}. These violent events have no counterpart in simulations
of solar convection performed with the ANTARES code \citep{Muthsam07}. Even though
they are intermittent in nature, they are common enough (cf.\ the front visible in
Fig.~\ref{Fig_tf_noHe} in the upper right corner of the simulation volume) to
be important statistically. They may even contribute to the shape of line profiles 
and provide an efficient source of heating up a chromosphere (cf.\ \citealt{Simon02}).

\subsection*{Resolution and computation of the radiative cooling rate $Q_{\rm rad}$}

Already for the case of solar granulation the most difficult quantity to resolve in equations
(\ref{eq.cont})--(\ref{eq.energy}) is the radiative heating and cooling rate $Q_{\rm rad}$
\citep{Nordlund91}. It directly influences the large, energy carrying scales by driving
the cooling (and reheating) of gas in the photosphere. Quantities directly related
to $Q_{\rm rad}$ such as the radiative flux $F_{\rm rad}$ and 
the superadiabatic temperature gradient $\nabla-\nabla_{\rm ad}$ or
the profiles of $T$ and $p$ are readily resolved at vertical grid spacings of
$\sim 12\dots 25$~km, but to obtain an equally smoothly resolved $Q_{\rm rad}$
requires a 10~times higher resolution \citep{Nordlund91}, even though 
$T$, $p$, and $F_{\rm rad}$ appear reasonably well converged at a resolution
of $\sim 12\dots 25$~km. Apparently, at that resolution the simulation runs
benefit from some averaging effects. We also recall that in regions where the diffusion 
approximation holds, we have $Q_{\rm rad} = -\mbox{\rm div} (K_{\rm rad} \nabla T)$
which involves a second derivative of $T$. This explains why resolution requirements
for $Q_{\rm rad}$ are higher than for mean structure quantities ($p$, $T$, \dots) and
their first derivatives ($F_{\rm rad}$, $\nabla$, \dots).

How well do we resolve $Q_{\rm rad}$ in A-type stars? Analysing our simulations
for the coarse grid, which at 200~km resolution is equivalent to a solar granulation
simulation with a resolution of $\sim 50$~km, the answer is: {\em at that grid spacing
not at all.} Near the surface $Q_{\rm rad}$ often has a triple-peak structure covered by 
about 15~grid points. At the same time on the fine grid there is only a double peak-structure
ranging just 5 such grid points, i.e.\ 20 points on the fine grid, whereas the maxima of these
peaks have less than $1/3$ of the size found
on the coarse grid. Physically, the double peak-structure originates from the radiative
cooling layer at the bottom of the photosphere (which is also present in RHD simulations 
of solar surface convection) and a heating layer just underneath it. The latter is created
by the partial ionisation of H~{\sc i} and the inefficiency of convective transport in the
superadiabatic layer coinciding with the entire surface convection zone (see
Fig.~\ref{Fig_gradients_He}). Once the simulation contains a fully developed surface
convection zone, the detailed structure of the peak becomes more complex, as 
heating and cooling layers are no longer confined to a thin zone (cf.\ Fig.~\ref{Fig_tf_He}
to Fig.~\ref{Fig_choc_b}). Since the coarse grid solution is discarded by ANTARES
at each time step for a domain where a fine grid solution is available, the wrong solution 
on the coarse grid has no impact on the development of the fine grid. But it nevertheless
implies a warning concerning simulations of mid A-type stars with lower effective 
resolution for $Q_{\rm rad}$ than the one presented by our fine grid solutions. We consider
even our current simulations on the boarder line of resolving the overall structure and 
magnitude of $Q_{\rm rad}$, because each of the very sharp peaks is represented by
only about 5~points. By comparison, at a vertical resolution of 16~km (corresponding
to about 60~km for our present case) a solar granulation simulation would have the
main (negative) peak covered by 15~points which resolve $Q_{\rm rad}$ smoothly 
everywhere except near its extremum \citep{Obertscheider07}. Thus, A stars require a 
3~to 4~times higher effective resolution at the stellar photosphere to properly 
resolve $Q_{\rm rad}$. This is also in agreement with the fact that $\nabla$ is 
steeper than in the Sun by just about that factor (Fig.~\ref{Fig_gradients_He}).

There is a second problem associated with $Q_{\rm rad}$. The radiative time 
scales in the photospheres of A stars are much shorter than in the Sun
\citep{Freytag95,Freytag04,Steffen07}, since the opacity at their surface is
lower while temperatures are higher. For optically thick layers 
the lower densities near their surface play a role, too. Hence, the time 
scale for relaxing a temperature perturbation of arbitrary optical thickness by
radiation \citep{Spiegel57}, 
\begin{equation}  \label{eq.t-rad}
   t_{\rm rad} = \frac{c_{\rm v}}{16 \kappa \sigma T^3}
           \left(1 - \frac{\kappa\rho}{k}\, \mbox{\rm arccot}\,\frac{\kappa\rho}{k} \right)^{-1},
\end{equation}
becomes smaller than the hydrodynamical time scales in a simulation. This includes 
the time scale of sound waves travelling a grid distance $h$, i.e., $t_{\rm sound} = h/c_{\rm s}$.
In Eq.~(\ref{eq.t-rad}), $c_{\rm v}$ 
is the specific heat at constant volume, $\kappa$ is the opacity, $\sigma$ is the 
Stefan-Boltzmann constant, and a perturbation of size $l$ with $k=2\pi / l$ is assumed.
For the optically thick case $t_{\rm rad}$ converges to the time scale of radiative diffusion,
$t_{\rm diff}=3(\kappa\rho/k)^2 c_{\rm v} / (16 \kappa \sigma T^3) \sim l^2 / \chi$, but
remains larger than zero for the optically thin case defined by $\kappa\rho / k \ll 1$
(using the specific heat at constant pressure, $c_{\rm p}$, in the definitions does not
change the argument). Comparing the time scales $t_{\rm sound}$ and $t_{\rm rad}$ for the
case $h=l$ reveals the shortest radiative relaxation time scales in the problem. It is important
to note that although for an A-type star the sound speed $c_{\rm s}$ is larger than for the 
Sun in the layer where $T=T_{\rm eff}$, $t_{\rm rad}$ becomes even smaller. For the Sun
at usual resolutions of $20\dots 30$~km for granulation simulations \citep{Kupka09} 
$t_{\rm sound} \lesssim t_{\rm rad}$, while for an A-type star with an equivalent
resolution $t_{\rm rad} \sim 0.01 \dots 0.1\, t_{\rm sound}$ for layers around the
optical surface \citep{Freytag95,Freytag04}.

As a consequence, RHD simulations of mid A-type stars, which use a purely explicit time 
integration method, are limited to time steps $\Delta t \leq t_{\rm rad}$. This was noticed
in \citet{Freytag04} and also in \citet{Freytag95}. Since ANTARES currently uses a purely
explicit time integration scheme for  Eqs.~(\ref{eq.cont})--(\ref{eq.energy}),
it is subject to the same restrictions. The most severe limitations implied by (\ref{eq.t-rad})
are found for optical depths $1 \lesssim \tau \lesssim 10$. There, $t_{\rm diff}$ is already
a useful approximation of $t_{\rm rad}$, whence $\Delta t \lesssim h^2/\chi$ as for the heat
equation. Consequently, a grid refinement by a factor of 4 in that region requires a 16 times
smaller time step. This explains the extremely small $\Delta t$ of 5~ms and 3.4~ms for
our 2D simulations with solar helium abundance and without helium, respectively.
We note that the coarse grid time steps of about 0.08~s and 0.05~s are comparable 
to the  $\lesssim 0.2$~s reported in \citet{Freytag04} --- these differences in $\Delta t$
show that the 2D simulations presented here require computational efforts just slightly smaller
than state-of-the-art 3D simulations at lower resolution $h$. But are such small $\Delta t$ 
unavoidable, if $Q_{\rm rad}$ is computed on the same mesh as the hydrodynamical variables? 
To check this we have computed the evolution time scale of the independent hydrodynamical 
variables $\rho$ and $e$. If we know both variables at the grid at a time step $n$, 
i.e., $\rho^{(n)}$ and $e^{(n)}$, for any integration method we require that
$\Delta t < C \rho^{(n)} (\partial \rho^{(n)} / \partial t)^{-1}$ and
$\Delta t < C e^{(n)} (\partial e^{(n)} / \partial t)^{-1}$, where $C$ is a constant less than 1, 
typically 0.1, to be able to predict the new state of the system at the time step $n+1$
(otherwise, even with fully implicit methods, iterative solvers may not converge, instabilities
can occur, etc.). If we take $C=0.1$, this is equivalent to requiring that at each grid cell the
solution should not change by more than 10\% during an integration step. Inspecting the
simulation at three subsequent time steps we can easily estimate the time derivative and
evaluate these inequalities. We have done this for the first time steps of the simulation and
for a number of time steps spread over the entire duration of the run for solar
helium abundance. To interpret the results we have also evaluated the convective flow time
scale $t_{\rm c} = h / \max(  ({\bf u}^2)^{1/2} )$ as well as $t_{\rm s} = h / \max c_{\rm s}$,
and finally $\Delta t < C e^{(n)} (Q_{\rm rad}^{(n)})^{-1}$ to see, if $Q_{\rm rad}$ operates on a 
shorter time scale than any of the mechanical terms in Eq.~(\ref{eq.energy}).
\begin{table}[!ht]
\caption{Time scales of the simulations with solar helium abundance. Grid 1 denotes the
   entire coarse grid, grid 2 the fine grid, grid 3 the coarse grid without the domain for which
   the fine grid is available. $C=0.1$ and $D=0.25$ in all calculations.}  \label{tab.timescales}
\begin{center}
\begin{tabular}{lrrr}
\hline\hline
time scale & grid 1 & grid 2 & grid 3 \\
\hline
\multicolumn{4}{c}{$t = 0.06$~s} \\
\hline
$D t_{\rm s}$            &        1.27  &       0.51  &         1.27  \\
$D t_{\rm c}$            & 18673.50  &  2204.89  &  36166.30  \\
$C \min(\rho/\rho_t)$ &        1.48  &     31.11  &         1.48  \\
$C \min(e/e_t)$        &        0.56   &      0.12  &         1.49  \\
$C \min(e/Q_{\rm rad})$ &   0.40   &      0.12  &       17.97  \\
\hline
\multicolumn{4}{c}{$t = 3.62$~s} \\
\hline
$D t_{\rm s}$            &        1.27  &       0.51  &         1.27  \\
$D t_{\rm c}$            &    371.56  &      23.12  &      587.73  \\
$C \min(\rho/\rho_t)$ &      25.60  &        7.60  &       25.60  \\
$C \min(e/e_t)$        &        7.33  &        2.91  &       20.33  \\
$C \min(e/Q_{\rm rad})$ &   0.84  &        2.15  &       64.36  \\
\hline
\multicolumn{4}{c}{$t = 11442.92$~s ($\sim 8.3\,t_{\rm sc}$)} \\
\hline
$D t_{\rm s}$            &        1.27  &       0.51  &         1.27  \\
$D t_{\rm c}$            &        1.86  &       0.49  &         8.05  \\
$C \min(\rho/\rho_t)$ &        0.80  &       0.53  &         6.30  \\
$C \min(e/e_t)$        &        0.81  &       0.53  &         7.35  \\
$C \min(e/Q_{\rm rad})$ &   0.40  &       1.12  &        13.59  \\
\hline
\end{tabular}
\end{center}
\label{default}
\end{table}

In Table~\ref{tab.timescales} we show the results for the third time step, for a time step
after initial relaxation through radiation, and for a time step during the generation of shock
fronts described in the previous subsection. Time steps between the second and the third
example show a rather continuous transition between these two. We note that only during the
first time steps the radiative heating and cooling ($e/Q_{\rm rad}$) constrains the temporal
evolution by providing the shortest time scale (even, if $C=D$). However, already after
a few seconds the temperature perturbations have been smoothed out to an extent that
sound speed and its associated time scale $t_{\rm s}$ set the restrictions for the time
evolution of the dynamical variables of the system. In the last time step shown supersonic 
velocities in the photosphere finally have led to $t_{\rm c} < t_{\rm s}$. We conclude that
except for the first few hundred simulation time steps totalling just a few seconds, which are 
subject to the assumed initial conditions and random perturbations applied to the latter,
the temporal evolution of the system is governed by the hydrodynamical time scales. This
holds even for $Q_{\rm rad}$ itself, which implies changes on the total energy on a similar
time scale as hydrodynamical processes (note that the overestimation of $Q_{\rm rad}$ on
the coarse grid leads to a slightly shorter time scale, but this part of the solution is discarded
by using the fine grid solution --- outside the fine grid domain the coarse grid solution imposes
no restrictions).

Thus, mathematically $Q_{\rm rad}$ is just a stiff term in a differential equation
making the whole problem at least in principle solvable on the evolution timescale of the
dynamical variables of the system by a properly designed implicit integration method. 
The restrictions in $\Delta t$ are solely due to high wave number components $k$ contained
in $Q_{\rm rad}$, which represent radiative transfer over one or a few grid cells. This transfer
indeed occurs on short time scales $t_{\rm rad}$, but does not govern the evolution of the
system itself, as it takes the much longer time $C e/Q_{\rm rad}$ to substantially change the 
energy content of a grid cell radiatively. A natural approach is to consider an operator splitting 
technique and integrate only $Q_{\rm rad}$ by an implicit method. This strategy is used also in 
simulations of convection in rotating spherical shells to model the lower part of the solar
convection zone \citep{Clune99}. But here the difficulty is that the coefficients in $Q_{\rm rad}$
are neither constant, nor linearly dependent on $(\rho,e)$. Plain subcycling for the
integration of $Q_{\rm rad}$ (multiple radiative transfer steps per hydrodynamical time step)
alone is inefficient, since $Q_{\rm rad}$ as a whole does not cause rapid changes to $e$, 
but only its high $k$ components evolve that way, while a consistent update of its coefficients
is non-trivial. Thus, we consider it more promising to analyse higher order methods with 
proper damping of small but rapidly evolving components or filtering methods for their 
suitability to accelerate RHD simulations of stellar convection at high resolution. This
would bring 3D simulations resolving the radiative heating and cooling at the surface
of mid A-type main sequence stars from a supercomputing application to the realm
of high performance department computers, as have been used in solar granulation
simulations with ANTARES (see \citealt{Muthsam07,Muthsam09}).

\section*{Conclusions and Outlook}

We have presented 2D RHD simulations performed with the ANTARES code for
a mid A-type star for two extreme cases of helium abundance, a solar one and
a helium free composition. The simulations differ from previous works by a higher
resolution and the application of high (5$^{\rm th}$) order advection schemes running
stably for these simulations without the need to introduce artificial diffusion (see
\citealt{Muthsam09} for further details). The quality of the advection scheme was
demonstrated by the need to introduce artificial damping of vertical velocities
over several sound-crossing time scales of the simulations to remove vertical
oscillations introduced by the initial conditions. 
After that phase oscillations driven by the flow itself are found
to reappear. The mean structure of the convection zone is close to a radiative
one, as found in previous RHD simulations in 2D and 3D. The most interesting
differences found between the two cases we have considered include an inversion
of the sign of the flux of kinetic energy as well as higher velocities and larger
flow structures, each of them observed for the case
with zero helium abundance. This can be understood in terms 
of the smaller extent of the convection zone due to the absence of partial
ionisation of helium and the lower mean molecular weight of a helium free mixture.
Since the evolution time scales of the surface convection zone are sufficiently
short, we expect these results to be robust. As a note of caution we point out 
the lack of strong vortices found for the zone of He~{\sc ii} ionisation for the case
with solar abundance, which is in contrast to the results reported in \citet{Freytag95}
and \citet{Freytag96}. From further simulation runs with different initial conditions
we have found that at least for the lower (He~{\sc ii} ionisation driven) convection
zone the simulations should be performed over very long times to ensure they
no longer depend on the initial perturbations (vortices seeded initially may just
remain while they may take a long time to grow on their own). As soon as the
surface convection zone is sufficiently developed large shock fronts are emitted
into the photosphere. These result from the interaction of vortex pairs with the
density inversion near the surface. We expect that a similar mechanism could work 
around strong downflow regions in 3D simulations as well. The large shock fronts
might affect spectral line profiles and provide a mechanism of heating chromospheres
in mid A-type stars. Finally, we have analysed the importance of resolution to 
properly compute the radiative heating and cooling rate $Q_{\rm rad}$ and have
shown that the time step restrictions on a hydrodynamical solver implied by
computing this quantity on a fine mesh are just those of a classical stiff term
in a differential equation. Properly designed implicit integration methods should
thus be able to accelerate RHD simulations of this class of stars.

This is most important since the two simulations shown here required
8.3~million time steps for the fine grid solutions resulting in a total of
$\sim 13,000$ CPU core hours (on POWER5 1.9 GHz CPUs). A 3D RHD simulation
at the same resolution $h$ is hence a supercomputing project (we recall that
this is a consequence of the $\Delta t \lesssim h^2/\chi$ dependence of an
explicit solver --- a larger $h$ as used in previous works would make the
simulation more affordable, but this should be traded in only, if $Q_{\rm rad}$ 
can be computed sufficiently accurately). Such a project is feasible with the
ANTARES code which has been successfully run with good scaling on up to
1024 CPU cores at the POWER6 machine of the RZG in Garching. However, 
this would still restrict us to very few simulation runs. We thus think it is important
to work on the implementation of proper integration methods, since this would
provide benefits also for simulations of other types of stars once high enough
resolution is demanded. Moreover, it would bring 3D RHD simulations of mid
A-type stars with a resolution $h$ as presented here into the realm 
of high performance department and university computers, which is already the
case for equivalent simulations of F-type to M-type main sequence stars. 

We are working on a version of ANTARES with open vertical boundary conditions
which avoids the reflection of waves and shock fronts. This will make the simulations
more suitable for the study of convection-pulsation interactions and increase the
stability of the code, since it avoids extreme conditions occurring when a front hits
the upper boundary. We intend to use long-term simulations with this new version
of ANTARES to compute synthetic spectra for comparisons with observations,
study convection-pulsation interactions, and probe non-local
models of convection, followed by 3D simulations as a final step.

\subsection*{Acknowledgments}
We thank Ch.~St\"utz for computing several 1D model atmospheres
with the LLmodels code which were used as initial conditions and also thank B. L\"ow-Baselli
for useful discussions on grid refinements. F.\,Kupka and J.\,Ballot are grateful to the MPI for 
Astrophysics and the RZG in Garching for granting access to the IBM POWER5 p575,
on which the numerical simulations presented in this paper have been performed. J.\,Ballot
acknowledges support through the ANR project Siroco and H.J.\,Muthsam is
grateful for support through the FWF project P18224-N13.

\end{document}